\documentstyle[12pt]{article}
\begin{document}
\vskip 2.6 truecm
\newcommand{\bb}{\begin{equation}}
\newcommand{\ee}{\end{equation}}
\newcommand{\pp}{\partial}
\newcommand{\lsim}{\buildrel <\over\sim}
\newcommand{\gsim}{\buildrel >\over\sim}

\begin{titlepage}

\title{Particle creation in a  colliding plane wave spacetime:
wave packet quantization} \author{  {\sc  Miquel Dorca  } \\
        {\small\em Grup de F\'\i sica Te\`orica  {\rm and} IFAE }\\
        {\small\em Universitat Aut\`onoma de Barcelona } \\
        {\small\em 08193 Bellaterra (Barcelona), Spain }\\
        { }\\
        {\sc Enric Verdaguer } \\
        {\small\em Departament de F\'\i sica Fonamental {\rm and} IFAE }\\
        {\small\em Universitat de Barcelona } \\
        {\small\em  Av. Diagonal, 647 } \\
        {\small\em 08028 Barcelona, Spain }  }

\date{\null}

\maketitle

\vskip -12cm\hskip 13cm UAB-FT-334
\vskip 12cm

\begin{abstract}

We use wave packet mode quantization to compute the creation of massless
scalar quantum particles in a colliding plane wave
spacetime. The background spacetime represents the collision of two
gravitational shock waves followed by trailing
gravitational radiation which focus into a Killing-Cauchy horizon. The use of
wave packet modes
simplifies the problem of mode propagation through the different spacetime
regions which was
previously studied with the use of monocromatic modes. It is found that the
number of particles
created in a given wave packet mode has a thermal spectrum with a temperature
which is inversely
proportional to the focusing time of the plane waves and which depends on the
mode trajectory.

\end{abstract}
\end{titlepage}


\section{Introduction}

Exact solutions representing the head on collision of two gravitational plane
waves are some of the simplest exact dynamical
spacetimes. They provide clear examples of highly non linear behavior in
general
relativity: when two plane waves collide the focusing effects of each exact
plane wave lead to
mutual fucusing. This is revealed by the formation of either a spacetime
singularity or of a
non-singular Killing-Cauchy horizon at the focusing points of the two waves
\cite{kah,gri,sze}.
They may also be useful to provide local models for processes that may be
taking place in our
universe as a result of gravitational waves produced in  black hole collisions
\cite{eat,fer-1},
the decay of cosmological inhomogeneous singularities \cite{yur-1}, or by
travelling waves in
strongly gravitating cosmic strings \cite{fin,vach}.

Quantum effects in such dynamical spacetimes must surely be important and one
expects particle production and vacuum
polarization when a quantum field is coupled to such background. Yurtsever
\cite{yur-1} was the first to study field
quantization on a colliding wave background, he considered the Kahn and
Penrose \cite{kah} solution which may be interpreted
as the collision of two impulsive plane waves. The solution which has
curvature singularities at the focusing points of the
plane waves allows for the definition of two physically meanigful vacuum
states: an ``in" vacuum state associated
to the flat space before the collision of the two plane waves and an ``out"
vacuum state related to the flat spacetime regions
behind the shock fronts also before the collision. The Bogoliubov coefficients
relating the ``in" and ``out" creation and
annihilation operators could be found only approximately in the
long-wavelenght limit. In this approximation the spectrum of
created particles is consistent with a thermal distribution.

In a recent paper \cite{dor}  we have considered the quantization of a
massless scalar field in the background of the collision
of two plane waves which form a non singular Killing-Cauchy horizon. This
solution
describes the collision of two gravitational shock waves followed by trailing
gravitational radiation \cite{hay}. The
interaction region of the two waves is locally isometric to a region inside
the event horizon of a
Schwarzschild black hole with the Killing-Cauchy horizon corresponding to the
event horizon
\cite{gri,cha-1,cha-2}. Two unambiguous and physically meaningful quantum
vacuum states may be
defined: an ``in" vacuum associated to the positive frequency mode solutions
in the flat region
before the collision of the waves  and an ``out" vacuum  related to the
positive frequency modes
defined through the two null vector fields in the Killing-Cauchy horizon. Such
state, which is
invariant under the symmetries associated to the horizon corresponds to the
unique {\it preferred
vacuum state} defined by Kay and Wald \cite{kay} in spacetimes with bifurcate
Killing  horizons.
It was found that the ``in" vacuum contains a number of ``out" particles which
is proportional to
the inverse of the frequency. In the long-wavelength limit the spectrum is
consistent with a
thermal spectrum at a temperature which is inversely proportional to the
focusing time of the
gravitational plane waves, in agreement with Yurtsever's result \cite{yur-1}.

Our result however is exact: although the Klein-Gordon equation in the
interaction region cannot be solved exactly the ``in"
modes become blueshifted towards the trailing points of the waves and can be
propagated through this region by using the
geometrical optics approximation. This is somewhat similar to the situation in
a Schwarzschild
black hole \cite{haw}.

In this paper we want to reconsider this problem in the light of wave packet
mode quantization instead of the monocromatic
modes we have used in ref. \cite{dor}. There are several reasons that, we
believe, justify this.
One of the reasons is well known: the use of monocromatic plane wave modes
leads to infinite
expressions for the total number of particles created of a given frequency,
whereas the number of
particles created in a given wave packet mode is finite \cite{haw}. The
second, and more
important for us here,  is that wave packet modes propagate in a simple way in
a spacetime,
such as ours, formed by
the matching of different regions. Since the singular free coordinates of
one of the regions differ from the next region, the propagation of modes which
are extended through all
space finds serious difficulties. In ref \cite{dor} this could be done because
of the blueshift
of the modes in certain regions and the use there of the geometrical optics
approximation which
ammounts to ray propagation. Since wave packets are localized in space and the
packet maximum
follows a well defined path in the spacetime, a natural approximation may be
taken by which the
propagation of the packets through the different regions becomes a simpler
problem.

Another reason is that the wave packet formalism localize the fenomena of
particle creation: the
one particle states defined with the wave packet modes have two labels, one
gives information on
the ``energy" of the particle and the other on its ``trajectory" (all within a
certain range of
values). Finally, we know that our colliding wave spacetime can be maximally
extended through the Killing-Cauchy horizon with the extended Schwarzschild
spacetime
\cite{hay,yur-2}. This is possible if one of the transversal coordinates of
the plane waves is
made cyclic. The resulting spacetime represents the collision of two plane
waves propagating in a
cylindrical universe and the creation of a black hole of mass proportional to
the strength (or
focusing time) of the plane waves. In a forthcoming paper we want to consider
particle creation
in the extended spacetime. Using the results of the present paper the
calculation will become
somewhat similar to that of stimulated emission by black holes \cite{wal-2};
the use of wave
packet modes has proved useful also in this case, see ref. \cite{mul}.

The plan of the paper is the following. In section 2 we introduce wave packet
mode quantization. In section 3 we briefly
review  the geometrical properties of our colliding wave spacetime with
special emphasis in the
coordinates which are appropriate in the different regions. In section 4 we
quantize a massless
scalar field on the colliding wave background and propagate the ``in" wave
packet modes through
the different regions. The advantage of wave packet modes is clearly seen in
subsection 4.4 when
the modes are propagated through the interaction region. The Bogoliubov
coefficients relating
packet creation and annihilation operators are derived in subsection 4.6, and
the creation of
particles is derived in subsection 4.7 where we also compare our results with
those of ref.
\cite{dor}.

\section{Wave packets}

Let us consider a complete orthonormal family, $\{f_{\omega}({ x})\}$, of
complex solutions of the Klein-Gordon equation for a
massless scalar field $\phi$ (i.e. $\Box\phi =0$), which contain only positive
frequencies with respect to a given timelike
Killing vector ${\partial}/{\partial}t$, i.e. such that  ${\cal
L}_{{\partial}/{\partial}t}\, f_{\omega}({ x})=-i\omega f_{\omega}({ x})$ with
$\omega >0$. Then it
is possible to define a  positive definite inner product between these
solutions. The label $\omega$ is continuous and stands
for the energy (or frequency) and ${ x}$ stands for the spacetime coordinates.
Since these solutions have a well defined value
for the energy (or frequency) we can call them {\it monocromatic modes}. It is
true that since the energy of the monocromatic
modes is well defined their space localization is completely uncertain in
concordance with the Heisenberg's uncertainty
principle.

Now we want to work with a complete and orthonormal set of modes that are
localized in space in some sense. To
achieve this we can make an adequate superposition, within a small energy
range, of continuous
$\omega$-labeled monocromatic modes in order to introduce an uncertainty in
the energy and
gain a certain information on spatial localization. We can do this as follows
\cite{haw,mul,mul'}, define  \bb {f}_{{\tilde\omega},n}({
x})={1\over\sqrt{\epsilon}}\int
_{{\tilde\omega}}^{{\tilde\omega}+\epsilon}d\omega\,{\mbox{\rm
e}}^{-in\omega}\, {f}_{{\omega}}({ x}),\label{eq:wp} \ee
where the new labels $\tilde\omega$ and $n$ are restricted to verify that
${\tilde\omega} /\epsilon\equiv j$ and $n\epsilon
/2{\pi}\equiv l$ are integers, $\epsilon$ being a small and positive
parameter. We will call this superposition a
{\it wave packet}. Here $n$ is a kind of Fourier label (the phase term
$\exp(-in\omega)$ has $l$ periodes in the interval $({\tilde\omega},\,
{\tilde\omega}+\epsilon)$), $\tilde\omega$ is the lower extremum of the
integration interval
and gives information on the energy of the wave packet. It can be shown easily
that a set of discrete (${\tilde\omega},n$)-labeled wave packets
$\{f_{{\tilde\omega},n}({ x} )\}$
is complete and orthonormal if the set of continuous $\omega$-labeled
monocromatic modes
$\{f_{\omega}({ x})\}$ is complete and orthonormal.

It is worth noticing that a general set of wave packets,
$\{f_{{\tilde\omega},n}({ x})\}$, given by
(\ref{eq:wp}), satisfy the following property,
\bb \sum _{j,\, l} \left|f_{{\tilde\omega},n}({ x})\right|^2=\int _0^\infty
d\omega\,
\left|f_{{\omega}}({ x})\right|^2, \label{eq:wpprop}  \ee
which follows from the
equality,
\bb \sum _{l=-\infty}^{\infty}\,{\mbox{\rm e}}^{{\pm}in\left(\omega -\omega
'\right)} = \epsilon\delta\left(\omega
-\omega '\right).\label{eq:prop}\ee
Note that the sums in (\ref{eq:wpprop}) and (\ref{eq:prop}) are over the
integer labels
$(j,\, l)$ which could also be used to label the modes, i.e.
$f_{j,l}(x)$, instead of the labels $(\tilde\omega,\, n)$ which we will use
throughout.

We can now see in what sense the wave packets (\ref{eq:wp}) are localized. Let
us consider the
generic wave packet (\ref{eq:wp}) in terms of the modulus and phase of
$f_{\omega}({ x})$, i.e. $f_{\omega}({
x})=\left|f_{\omega}({ x})\right|\, {\mbox{\rm e}}^{i{\theta}_{\omega}({
x})}$, as:
 \bb {f}_{{\tilde\omega},n}({
x})={1\over\sqrt{\epsilon}}\int
_{{\tilde\omega}}^{{\tilde\omega}+\epsilon}d\omega\,{\mbox{\rm e}}^{-in\omega
+i{\theta}_{\omega}({ x})}\, \left|f_{\omega}({ x})\right|.\label{eq:wpf} \ee
If the interval
of integration in (\ref{eq:wp}) is small enough so that
$\left|f_{\omega}({ x})\right|$ can be taken as approximatly constant (this
will be
true in general provided that $\left|f_{\omega}({ x})\right|$ has no
singularities in the interval of integration),
the $\left|f_{\omega}({ x})\right|$ can be factorized out and we  have an
integral over the phase
only. Now if the integrand's phase, i.e.
${\Theta}_{\omega}({ x})\equiv -n\omega +{\theta}_{\omega}({ x})$, oscillates
rapidly over the
range of integration (at least when $n\omega$ is big enough, and this is true
for
$l>\! >1$) then the integral roughly vanishes except at the stationary  phase
points, that is when  \bb \left.{{\partial}{\Theta}_{\omega}({ x})\over
{\partial}\omega}\right|_{\omega={\tilde\omega}} =0,\label{eq:wpt} \ee and we
can use the stationary phase
method \cite{eck} to evaluate (\ref{eq:wpf}). Note that in the equation
(\ref{eq:wpt}) we have
set $\omega =\tilde\omega$, after derivation, which is accurate provided the
range of energy
superposition is small enough. Equation (\ref{eq:wpt}) stablishes a relation
between the
labels ($\tilde\omega$, $n$) of the wave packet and the spacetime coordinates
${ x}$ and so it
represents a three dimensional hypersurface. It determines the geometric locus
of
the spacetime points which give the main contribution to the integral
(\ref{eq:wp}), this will be a spacetime region labeled by
$\tilde\omega$ and $n$ and we will identify it as the {\it wave packet
trajectory}, because outside this region the integral
(\ref{eq:wp}) roughly vanishes. It is worth noticing that if the monocromatic
modes are labeled with some other
continuous parameters, besides the energy, we can construct double or triple
wave packets by
superposition of monocromatic modes in a small range of these parameters. In
this case we obtain
surfaces or curves as the trajectories of the double or triple wave packet,
respectively.

Finally we can make the following remark about the uncertainty of the energy
and position of a
wave packet. When we construct a wave packet we take a superposition of
Klein-Gordon solutions with a well defined energy $\omega$ (the monocromatic
modes) over a small
interval $(\omega,\, \omega +\epsilon)$ , and so the uncertainty in the energy
of the wave packet
is $\Delta\omega\simeq\epsilon$. To know the uncertainty in the position of
the wave packet let
us assume the following simple form for the monocromatic modes $f_{\omega}({
x})={\rm
e}^{-i\omega t}\, F_{\omega}({x^j})$ (where $x^j,\;\; j=1,2,3$ are space
coordinates) which
verify ${\cal L}_{{\partial}/{\partial}t}\, f_{\omega}({ x})=-i\omega
f_{\omega}({ x})$. Then with the use of the
stationary phase method we easily see that the wave packets  (\ref{eq:wpf})
are peaked arround
values of time given by $t=-2{\pi}l{\epsilon}^{-1}$, with width
$2{\pi}{\epsilon}^{-1}$ ($l$ is an
integer) and so the time uncertainty is $\Delta t\simeq
2{\pi}{\epsilon}^{-1}$, which reflects
Heisenberg's uncertainty principle $\Delta\omega\,\Delta t\simeq 2{\pi}$.

\section{Colliding plane waves geometry}

We will work in a spacetime that describes the head on collision of two
linearly polarized gravitational plane waves
propagating in the $z$-direction. This spacetime has four regions (see Fig.
1): a flat region (or region IV) at the past, before
the arrival of the waves, two plane wave regions (regions II and III) and an
interaction region (region I) where the waves
collide and interact nonlinearly. The geometry of these regions is given by
the following four metrics, in coordinates which
are adapted to the Killing vectors ${\partial}_x$ and ${\partial}_y$ of the
spacetime by (see ref.
\cite{gri} and references therein),
$$ds_{\rm
I}^2=4{L_1L_2}\,{\left[{1+\sin\left({u+v}\right)}\right]}^{2}dudv-{1-\sin%
\left({u+v}\right) \over
1+\sin\left({u+v}\right)}d{x}^{2}-$$ \bb
-{\left[{1+\sin\left({u+v}\right)}\right]}^{2}{{\cos}^{2}\left({u-v}%
\right)}d{y}^{2}{}^{} , \label{eq:MI}\ee
\bb\begin{array}{c}\displaystyle
{ds^2}_{\rm
II}=4{L_1L_2}\,{\left[{1+\sin\left({u}\right)}\right]}^{2}dudv-{1-\sin\left({u}%
\right)
\over 1+\sin\left({u}\right)}d{x}^{2}\\
\\
-{\left[{1+\sin\left({u}\right)}\right]}^{2}{{\cos}^{2}\left({u}%
\right)}d{y}^{2}{}^{},\end{array}
\label{eq:MII}\ee
\bb\begin{array}{c}\displaystyle
{ds^2}_{\rm
III}=4{L_1L_2}\,{\left[{1+\sin\left({v}\right)}\right]}^{2}dudv-{1-\sin%
\left({v}\right)
\over
1+\sin\left({v}\right)}d{x}^{2}\\
\\
-{\left[{1+\sin\left({v}\right)}\right]}^{2}{{\cos}^{2}\left({v}%
\right)}d{y}^{2}{}^{},\end{array}
\label{eq:MIII}\ee
\bb{ds^2}_{\rm IV}=4{L_1L_2}\,dudv-d{x}^{2}-d{y}^{2}, \label{eq:MIV}\ee
where $u$ and $v$ are two dimensionless null coordinates ($v+u$ is a time
coordinate and $v-u$ a space
coordinate) and $L_1$, $L_2$ are two arbitrary positive lenght parameters,
which represent the inverse of the strength
(focusing time) of the waves. The boundaries of these four regions are:
$\{u=0$, $v\leq 0\}$
between regions IV and II, $\{v=0$, $u\leq 0\}$ between regions IV and III,
$\{v=0$, $0\leq
u<{\pi}/2\}$ between regions II and I and $\{u=0$, $0\leq v<{\pi}/2\}$ between
regions III and I.

At the surfaces $u={\pi}/2$ and $v={\pi}/2$ on regions II and III,
respectively, the determinants of the respective
metrics vanish, this marks the focusing points of the waves and a coordinate
singularity. This singularity can be
avoided with the use of appropriate coordinates (harmonic coordinates) in
which the causal structure of these spacetime regions
is well posed. In these coordinates the surfaces $u={\pi}/2$ and $v={\pi}/2$
become spacetime lines, see\cite{dor,hay} for
details.

Region I
(in Fig. 1, this is the triangle bounded by the lines $\{v=0$, $0\leq
u<{\pi}/2\}$, $\{u=0$, $0\leq v<{\pi}/2\}$ and
$u+v={\pi}/2$ ) is locally isometric to a region of the interior of the
Schwarzschild metric. This is easily seen with the coordinate transformation,
$$t=x,\;\; r=M\left[{1+\sin (u+v)}\right],$$
$$\varphi =1+y/M,\;\; \theta = {\pi}/2-(u-v)
$$
where we have defined $M=\sqrt{L_1 L_2}$. The metric (\ref{eq:MI}) becomes
$$ ds^2=\left({{2M \over r}-1}\right)^{-1}d{r}^{2}-\left({{2M
\over r}-1}\right)d{t}^{2}-{r}^{2}\left({d{\theta }^{2}+{{\sin}^{2}\theta
}d{\varphi }^{2}}\right), $$
which is the interior of the Schwarzshild metric. The surface $u+v={\pi}/2$
corresponds to the black hole event horizon. The boundary $v=0$ corresponds to
$r=M(1+\cos\theta)$ and $u=0$ corresponds to $r=M(1-\cos\theta)$. These are the
boundaries of the plane waves, these boundaries join at
$r=M$ (spacetime point of the collision) and also at the surface $u+v={\pi}/2$
at $\theta
=0$ and $\theta ={\pi}$. This region of the Schwarzschild interior does not
include the
singularity $r=0$ and thus the interaction region has no curvature
sin\-gu\-la\-ri\-ties. The above local isometry is not global however, the
coordinates $\theta$ and $\phi$ are cyclic in the black hole case but in the
plane wave case, $-\infty <y<\infty$ and $-\infty <v-u<\infty$.

As in the the Schwarzschild case it is
convenient to introduce a set of Kruskal-Szekeres like coordinates to describe
the interaction region, because the $(u,\, v,\, x,\, y)$ coordinates become
singular at the horizon. These coordinates will
play  an important role in the quantization of the field.  First we introduce
dimensionless time and space
coordinates $(\xi ,\eta)$
\bb \xi =u+v,\;\; \eta =v-u
,\label{eq:XIETA}\ee
with the range
$0\leq\xi <{\pi}/2,\;\; -{\pi}/2\leq\eta <{\pi}/2$
(we shall later see that in these coordinates the Klein-Gordon
equation can be separated). Then we introduce a new time coordinate ${\xi}^*$
related to the dimensionless time coordinate $\xi$ by
\bb {\xi
}^{{}^*}=2M\ln\left({{1+\sin\xi  \over 2{{\cos}^2\xi }}}\right)-M\left({\sin\xi
-1}\right) ,\label{eq:XI*}\ee
and a new set of null coordinates
\bb {\tilde U} ={\xi}^*-x,\;\;
{\tilde V}={\xi}^*+x. \label{eq:UVtilde}\ee
Note that the transversal coordinate $x$ appears in the coordinate
transformation because it behaves badly at the horizon.
Finally, we define, \bb U'=-2M{\exp}\left({-{{\tilde U}\over 4M}}\right)\leq
0,\;\;
V'=-2M{\exp}\left({-{{\tilde V}\over 4M}}\right)\leq 0,\label{eq:U'V'}\ee
and the metric in the interaction region (\ref{eq:MI}) reads,
$$  d{s}^{2}_I={{2{\exp}\left[(1-\sin\xi )/2\right] \over
\left({1+\sin\xi }\right)}}dU'dV'-{M}^{2}{\left({1+\sin\xi
}\right)}^{2}d{\eta }^{2}-$$
$$-{\left({1+\sin\xi }\right)}^{2}{\cos^{2}\eta }d{y}^{2} ,\label{eq:HM}$$
with
\bb U'V'=8{M}^{2}{{{\cos}^{2}\xi } \over 1+\sin\xi }{\exp}\left({{\sin\xi -1
\over 2}}\right) ,\label{eq:U'.V'}\ee
\bb { U' \over V'}={\exp}\left({{x \over 2M}}\right) ,\label{eq:U'/V'}\ee
The curves $\xi =const.$ and $x=const.$ are, respectively, hyperbolae
and straight lines through the origin of coordinates $(U'=V'=0)$, see Fig. 2.
The Schwarzshild horizon (which is a Killing-Cauchy horizon for the spacetime)
corresponds to the limit of the
hyperbolae when $\xi\rightarrow {\pi}/2$ i.e. the ``roofs" $ V'=0$ or $U'=0$.
Notice that the
problem with the transversal coordinate $x$ at the horizon is that all the
lines
$x=const.$ go through the origin of the $(U',V')$ coordinates, so that all the
range of $x$ collapses into the point $V'=U'=0$, whereas the lines $U'=0$ and
$V'=0$ represent $x=-\infty$ and $x=\infty$ respectively. One should  recall
that we have not represented the coordinate $x$ in our picture of the collision
(Fig. 1) in which only the $(u,v)$ coordinates are shown, $x$ is a transversal
coordinate perpendicular to the propagation and adapted to
the Killing vector ${\partial}_x$.

To understand the global geometry of the spacetime one needs a tridimensional
picture
where the boundary surfaces
between the different regions have to be written in terms of appropriate
nonsingular coordinates
adapted to each region. See \cite{dor,hay} for details.

It is worth noticing that most of the plane wave
collisions produce true curvature singularities \cite{gri}, but the spacetime
described above is an example of a collision where
the curvature singularity has been substituted by a
Killing-Cauchy horizon (i.e. the surface $U'V'=0$ in the Kruskal-Szekeres like
coordinates).

\section{Wave packets in the collision of two gravitational plane waves}

\subsection{Monocromatic mode quantization (summary)}

In ref. \cite{dor} the quantization of a massless scalar field $\phi$ was
considered
in the spacetime described above representing  the head on collision of two
gravitational
plane waves. Let us now briefly summarize this quantization sheme. We
introduce a massless scalar
field $\phi$ on the colliding spacetime background, which satisfies the
Klein-Gordon equation,
\bb \Box\phi =0, \label{eq:KG'}\ee where $\Box\phi
=(-g)^{-1/2}\left((-g)^{1/2}g^{\mu\nu}\phi _{,\nu}\right)_{,\mu}$, and $g$ is
the determinant of the metric. It has the
generic plane wave solution \bb \phi ={1 \over \sqrt
{FG}}f(u,v){\mbox{\rm e}}^{i{k}_{x}x+i{k}_{y}y},\label{eq:SEP} \ee
where the labels $k_x$ and $k_y$ are two separation constants which are
interpreted physically as
the momenta in the directions given by the Killing vectors ${\partial}_x$ and
${\partial}_y$, which generate the
plane symmetry of the whole spacetime. Introducing (\ref{eq:SEP}) in
(\ref{eq:KG'}) one obtains,
\bb { f}_{,uv}-\left[{{{\left({\sqrt {FG}}\right)}_{,uv} \over \sqrt
{FG}}-{{\mbox{\rm e}}^{-N} \over 4}\left({{{{{k}_{x}}}^{2} \over
{F}^{2}}+{{{{k}_{y}}}^{2} \over {G}^{2}}}\right)}\right]f=0, \label{eq:KG} \ee
where the coefficients $F$, $G$ and ${\mbox{\rm e}}^{-N}$, come from the
adoption of a generic metric for all colliding
plane wave spacetime, adapted to ${\partial}_x$ and ${\partial}_y$, i.e.
\bb d{s}^{2}={\mbox{\rm
e}}^{-N(u,v)}dudv-{F}^{2}(u,v)d{x}^{2}-{G}^{2}(u,v)d{y}^{2}. \label{eq:GM} \ee

One
defines an initial vacuum state constructed with a complete orthonormal set of
solutions of the
Klein-Gordon equation (\ref{eq:KG'}) which are defined in the flat IV region,
before the arrival
of the plane waves, to be of positive frequency with respect to the timelike
Killing vector
${\partial}_{u+v}$; these are the ``in" monocromatic modes. Then one
propagates these modes throughout the
spacetime up to the horizon of region I (solving the appropriate boundary
conditions imposed by
the different classes of solutions for the Klein-Gordon equation,
(\ref{eq:KG}), in the different
regions of the spacetime). It is possible to define another ``natural" vacuum
state on the
horizon of region I from a complete and orthonormal set of solutions of the
Klein-Gordon equation
(\ref{eq:KG'}), the ``out" monocromatic modes, which are positive frequency
solutions with
respect to the vectors ${\partial}/{\partial}U'$ and
${\partial}/{\partial}V'$, which are two null Killing fields over the horizon.

Comparing the propagated ``in" monocromatic modes and the ``out" monocromatic
modes on the horizon
via a Bogoliubov transformation, lead us to show that there is spontaneous
creation of particles
in this spacetime with a spectrum of ``out" particles given by the formula
\cite{dor},
\bb \langle 0,\,{\mbox{\rm in}}|N_{\omega _-}^{\rm out}|0,\,{\mbox{\rm
in}}\rangle =
{(2M)^3\over {\pi}}\,{\delta _{m,-{\tilde m}}\over (8{\pi}M\omega _-) }\,\int
d{\hat k}_-\int
dk\,{\left|{C_{l'}}\right|^2\over{\mbox{\rm e}}^{k}-1}
. \label{eq:CSPT} \ee
Here $N_{\omega _-}^{\rm out}$ is the usual number operator for ``out"
particles, $\omega _-$,
$l'$ and $m'$ are labels of the  ``out" modes ($\omega _-$ is the energy
label), ${\hat k}_-$, $k$
and $m$ are dimensionless labels of the ``in" modes and $(2M)^3/{\pi}
|C_{l'}|^2$ is a geometric
factor whose coefficient, $|C_{l'}|$, depends on  $l'$, $m'$, ${\hat k}_-$ and
$k$. This
spectrum is inversely proportional to the inverse of the energy of the ``out"
particles which
are produced and it is consistent, in the long wavelenght limit, i.e.
$8{\pi}M\omega_-<\! <1$, with a
thermal espectrum with temperature $T=(8{\pi}M)^{-1}$. The temperature is
inversely proportional to
the focusing time of the plane waves, given by the parameter
$M=\sqrt{L_1L_2}$, and $(4M)^{-1}$
is also the surface gravity of the horizon. Note that the spectrum contains a
logarithmic
divergence, this divergence appears because the spectrum which has units of
$({\rm length})^3$,
given by the factor $(2M)^3$,  describes the total number of  particles
created in the whole
spacetime volume. As it is well known \cite{bir} this is characteristic of the
use of continuous
labeled modes and can be avoided using wave packets.

The propagation of the ``in" monocromatic modes throughout the spacetime is a
difficult task. This is due, essentially, to the
fact that the monocromatic modes are defined in the whole spacetime and there
is not a single singularity free coordinate
chart for all the four spacetime regions. Basically the problem is that the
matching has to be
done through all the spacetime points of the boundaries. Due to their spatial
localization wave
packet modes will be more easily matched. The matching will be done
approximately on a
single boundary point for each wave packet.

In what follows we introduce the wave packet  formalism in the colliding
spacetime background. As the first step
we construct a complete set of ``in" wave packets, from a superposition of
positive frequency monocromatic ``in"
modes, in the four spacetime regions. Then we will propagate these ``in" wave
packets throughout the spacetime
up to the horizon, with appropriate matching conditions. These matching
conditions will ensure that the
trajectory of the ``in" wave packet will be smoothly connected through the
boundaries between the  different regions. Next
we will contruct a complete set of ``out" wave packets on the surface of the
horizon. Finally, we will relate the ``in" and
``out" wave packets via a Bogoliubov transformation and compute creation of
particles in this formalism.

\subsection{Flat region (region IV)}

In the flat region the complete set of ``in"-modes is,
\bb {u}_{k_x k_yk_-}^{{\rm (IV)}}(u,v,x,y)={1\over \sqrt{2{{k}}_- {(2{\pi})}^3
}
}\,{\mbox{\rm e}}^{-ik_- v' -ik_+ u' +ik_x x+ik_y y} ,\label{eq:m-iv}\ee
where the labels $k_x$, $k_y$ and $k_-$ are independent separation constants
for the Klein-Gordon equation
(\ref{eq:KG'})
and where $u'$, $v'$ are two dimensional null coordinates related to the
dimensionless null coordinates $u$, $v$ by:
$v'=2L_2 v$, $u'=2L_1u$. The label $k_+$  is determined by the relation,
\bb 4k_+k_-=k_x^2+k_y^2.\label{eq:r-iv}\ee
It was shown in ref \cite{dor} that these modes are well normalized on the
hypersurface $\{u=0,\, v\leq 0\}\cup\{v=0,\, u\leq 0\}$.

The labels $k_x$ and $k_-$ are continuous but $k_y$ is discrete if we take a
cyclic spacetime
in the $y$-direction (this is not necessary but it is convenient if we want to
maximally extend the collidig wave spacetime
later on). We identify $k_y$ with $m/M$ where $m$ is an integer.  Our aim now
is to discretize the continuous
labels $k_x$ and $k_-$ by constructing a {\it double wave packet} as,
\bb {u}_{{\tilde k}_x,n, k_y,{\tilde k}_-,q}^{{\rm
(IV)}}(u,v,x,y)={1\over\sqrt{\epsilon\delta}}\int _{{\tilde
k}_x}^{{\tilde k}_x+\epsilon}dk_x\int _{{\tilde
k}_-}^{{\tilde k}_-+\delta}dk_-{{\mbox{\rm e}}^{-iqk_- -ink_x }\over
\sqrt{2{{k}}_- {(2{\pi})}^3 }
}\,{\mbox{\rm e}}^{-ik_- v' -ik_+ u' +ik_x x+ik_y y}.\label{eq:p-iv}\ee
Note that since we are only integrating over positive frequency monocromatic
modes the vacuum
associated with the wave  packets is the same as that defined by the
monocromatic modes
(\ref{eq:m-iv}).

The integrand's phase is
\bb \Theta ^{{\rm (IV)}}=-nk_x-qk_--k_-v'-\left({k_x^2+k_y^2\over
4k_-}\right)u'+k_xx+k_yy,\label{eq:f-iv}\ee
where the relation (\ref{eq:r-iv}) has been used. The trajectory of the double
wave packet, in the sense given by the
stationary phase method, i.e. ${\partial}\Theta ^{{\rm
(IV)}}/{\partial}k_x={\partial}\Theta ^{{\rm (IV)}}/{\partial}k_-=0$, is given
by,
\bb
x^{{\rm (IV)}}=n+{{\tilde k}_x\over {\tilde k}_-}L_1u,\label{eq:t1-iv}\ee \bb
v'^{{\rm (IV)}}=-q+{{\tilde
k}_x^2+{\tilde k}_y^2\over 2{\tilde k}_-^2}L_1u,\label{eq:t2-iv}\ee which
define a null geodesic in the flat region. That is,
this double wave packet moves on a null trajectory.

\subsection{Plane wave region (region II)}

The complete set of solutions of the Klein-Gordon equation (\ref{eq:KG'}) in
region II is easily found (see ref. \cite{dor})
and is given by
 \bb {u}_{k_x k_yk_-}^{{\rm (II)}}(u,v,x,y)={1\over \sqrt{2{{k}}_-
{(2{\pi})}^3 } }{1\over\cos
u}\,\exp\left({-{iL_1\over 2k_-}\left(f(u)k_x^2+g(u)k_y^2\right)-ik_- v'+ik_x
x+ik_y y}\right) ,\label{eq:m-ii}\ee where the
two fuctions $f(u)$ and $g(u)$ are \bb f(u)={{{\left({1+\sin u}\right)}^{2}
\over
2\cos u}\left({9-\sin u}\right)+{15 \over 2}\cos u-{15 \over
2}u-12},\;\;\;
g(u)=\tan u.\label{eq:fg}\ee
The labels $k_x$, $k_y$ and $k_-$ have the same meaning that in the flat
region IV because this expression for
the solutions of the Klein-Gordon equation in region II match smootly (i.e. in
a continuous and diferentiable way) with
the respective solutions (\ref{eq:m-iv}) on the boundary between regions II
and IV, i.e. $\{u=0$,
$v\leq 0\}$.

In analogy with the previous case, we now construct a double wave packet
$${u}_{{\tilde k}_x,n, k_y,{\tilde k}_-,q}^{{\rm
(II)}}(u,v,x,y)={1\over\sqrt{\epsilon\delta}}\int _{{\tilde
k}_x}^{{\tilde k}_x+\epsilon}dk_x\int _{{\tilde
k}_-}^{{\tilde k}_-+\delta}dk_-{{\mbox{\rm e}}^{-iqk_- -ink_x }\over
\sqrt{2{{k}}_- {(2{\pi})}^3 }
}$$
\bb \times {1\over\cos u}\,\exp\left({-i{L_1\over
2k_-}\left(f(u)k_x^2+g(u)k_y^2\right)-ik_-
v'+ik_x x+ik_y y}\right).\label{eq:p-ii}\ee
The integrand's phase is given by,
\bb \Theta ^{{\rm
(II)}}=-nk_x-qk_--k_-v'-\left({f(u)k_x^2+g(u)k_y^2}\right){L_1\over
2k_-}+k_xx+k_yy,\label{eq:f-ii}\ee
and the trajectory of the double wave packet, i.e. ${\partial}\Theta
^{{\rm (II)}}/{\partial}k_x={\partial}\Theta ^{{\rm (II)}}/{\partial}k_-=0$,
is \bb x^{{\rm (II)}}=n+{{\tilde k}_x\over
{\tilde k}_-}L_1f(u),\label{eq:t1-ii}\ee \bb v'^{{\rm
(II)}}=-q+\left({f(u){\tilde
k}_x^2+g(u){\tilde k}_y^2}\right){L_1\over 2{\tilde
k}_-^2},\label{eq:t2-ii}\ee which represents
a null geodesic travelling throughout the plane wave region II. Note that this
trajectory matches
smoothly (i.e. in a continuous and diferentiable way) with the trajectory of
the  wave packet in
region IV, i.e.  (\ref{eq:t1-iv}), (\ref{eq:t2-iv}), on the boundary  $\{u=0$,
$v\leq 0\}$. Since
$f(0)=g(0)=0$ and $df/du(0)=dg/du(0)=0$ the following matching conditions are
satisfied:
$$x^{{\rm (IV)}}(u=0)=x^{{\rm (II)}}(u=0);\;\; v'^{{\rm (IV)}}(u=0)=v'^{{\rm
(II)}}(u=0),$$
$$\left.{dx^{{\rm (IV)}}\over du}\right|_{u=0}=\left.{dx^{{\rm (II)}}\over
du}\right|_{u=0} ;\;\;
\left.{dv'^{{\rm (IV)}}\over du}\right|_{u=0}=\left.{dv'^{{\rm (II)}}\over
du}\right|_{u=0} .$$
This is not surprising because the monocromatic modes which we have used to
define the double
wave packets in the two regions, match smoothly on the boundary $\{u=0,$
$v\leq 0\}$.

Before going further, we can extract some information on the meaning of the
new discrete labels
$n$ and $q$. First, note that at the boundary between regions II and IV,
$\{u=0,$ $v\leq 0\}$, we
have from (\ref{eq:t1-iv}), (\ref{eq:t2-iv})  and (\ref{eq:t1-ii}),
(\ref{eq:t2-ii}),
$$x_0\equiv x^{{\rm (IV)}}(0)=x^{{\rm (II)}}(0)=n,$$
$$v'_0\equiv v'^{{\rm (IV)}}(0)=v'^{{\rm (II)}}(0)=-q,$$
which means that a wave packet labeled by $n$ and $q$ has a peak on the null
geodesic which crosses the flat region
into the plane wave region at the spacetime coordinates $x_0=n$ and $v'_0=-q$.
Note that for wave packets in region III we can repeat the same discussion.

\subsection{Interaction region (region I)}

In this region mode propagation is  more difficult; instead of relaying on the
calculations
given in ref. \cite{dor}, we will start the discussion from the beginning.
First, let us consider
that the Klein-Gordon equation (\ref{eq:KG'}) in this region can be separed by
taking  \bb \phi
(\xi ,\eta ,x,y)=e^{ik_x x+ik_y y}\,\psi _{\alpha k_x}(\xi)\,\varphi _{\alpha
k_y}(\eta),  \ee
and reduced to equations for $\psi _{\alpha k_x}(\xi)$ and $\varphi _{\alpha
k_y}(\eta)$. The
coordinates $\xi$, $\eta$ are related to the usual null coordinates $u$, $v$,
by
(\ref{eq:XIETA}), and the two new equations read, \bb {\psi}_{,\xi
\xi}-(\tan\xi
){\psi}_{,\xi }+\left( \alpha +{\hat k_x^{2} \over 4}{{(1+\sin\xi )}^4 \over
{\cos}^2 \xi}\right)\psi = 0, \label{eq:e1}\ee
\bb {\varphi }_{,\eta \eta}-(\tan\eta ){\varphi}_{,\eta }+\left( \alpha
-{\hat k_y^{2} \over 4}{1 \over {\cos}^2 \eta}\right)\varphi = 0  ,
\label{eq:e2} \ee
where $\alpha$ is a dimensionless separation constant and $k_x$, $k_y$ are the
same labels as in regions IV or II. We use the
notation ${\hat a}=2Ma$, therefore ${\hat k}_x$, ${\hat k}_y$ are
dimensionless parameters.

These differential equations have singular points at $\xi ={\pi}/2$, i.e. on
the horizon of region I, and at $\eta ={\pm}{\pi}/2$
respectively. To avoid them we can perform the following change of variables:
$ d\xi ^*=M(1+\sin\xi)^2 \cos^{-1}\xi\, d\xi$,
$ d\eta ^*=M \cos^{-1}\eta\, d\eta$.
With appropriate integration constants, $\xi ^*$ reduces to
(\ref{eq:XI*}) and $\eta ^*$ to:
\bb \eta ^*=M\,{\mbox{\rm ln}}\left({1+\sin\eta\over\cos \eta}\right).
\label{eq:eta*} \ee
We can still introduce a new function $\gamma _{\alpha k_x}(\xi)$ instead of
$\psi _{\alpha k_x}(\xi)$ by,
$ \gamma =(1/2)(1+\sin\xi)\psi$,
and we arrive at the following equations for $\gamma _{\alpha k_x}(\xi)$ and
$\varphi _{\alpha k_y}(\eta)$,
\bb {\gamma }_{,{\xi}^* {\xi}^*}+{\left[k_x^2 +{{\cos}^2\xi\over
M^2(1+\sin\xi )^4}\left(\alpha +{2\sin\xi\over 1+\sin\xi}\right)\right]}\gamma
=0,\label{eq:e'1}  \ee
\bb {\varphi }_{,{\eta}^* {\eta}^*}+{\left[-k_y^2 +{{\cos}^2\eta\over
M^2}\, \alpha\right]}\varphi
=0,\label{eq:e'2} \ee
which have no singular points in the regions of interest.

Recall that we are looking for a solution of the scalar field $\phi$ in the
interaction region,
restricted to satisfy certain boundary conditions imposed by the wave packets
travelling from
regions II (or III) into region I on the boundary $\{v=0,$ $0\leq u<{\pi}/2\}$
(or $\{u=0,$ $0\leq
v<{\pi}/2\}$) Thus the general solution for $\phi$ in terms of the new
functions
${\gamma}_{\alpha { k}_{x}}(\xi)$, ${\varphi}_{\alpha { k}_{y}}(\eta )$ is
given by, \bb \phi
(\xi ,\eta ,x,y) ={\mbox{\rm e}}^{ik_x x+ik_y y}\,\sum _{\alpha } C_{\alpha}\,
{{\gamma}_{\alpha
{k}_{x}}(\xi)\over 1+\sin\xi}\,{\varphi}_{\alpha { k}_{y}}(\eta )
,\label{eq:lin}   \ee
where the coefficients $C_{\alpha}$ depend on ${\alpha}$ and the separation
constants
used to label the monocromatic modes in region II, i.e. $k_x$, $k_y$ and $k_-$.
Now we will try to obtain all possible information on the coefficients
$C_{\alpha}$ in (\ref{eq:lin}). We know that expression (\ref{eq:lin}) gives
the general
solution (once $k_x$ and $k_y$ are fixed) for a massless scalar field in the
interaction
region, and the coefficients $C_{\alpha}$ have to be such that the appropriate
boundary conditions are satisfied. If we want to match the monocromatic modes
defined in region I
with those of region II (\ref{eq:m-ii}) (or III) we have to perform a sum over
all possible
values of $\alpha$ in (\ref{eq:lin}), this was one of the main difficulties in
ref. \cite{dor}. In
the following discussion we will show that when we match the  spatially
localized wave packets
travelling from region II (or III) into region I (\ref{eq:p-ii}) with wave
packets defined using
the general solution (\ref{eq:lin}), there is only one coefficient
$C_{\alpha}$ which carries the
main contribution. That is, we will find a single value of $\alpha$ for which
the infinite linear
combination (\ref{eq:lin}) can be approximated by a single term. Furthermore,
we will be able to
find the value of the coefficient $C_{\alpha}$ for that particular value of
$\alpha$.

We will proceed as follows. First, we will find an approximate solution of
equations (\ref{eq:e'1}) and
(\ref{eq:e'2}) for $\gamma _{\alpha k_x}(\xi)$ and
$\varphi _{\alpha k_y}(\eta)$ respectively, near the boundary of region I with
region II which will lead to an
approximate solution of (\ref{eq:lin}) near this boundary. With this solution
we will construct
a wave packet in analogy with (\ref{eq:p-iv}) or (\ref{eq:p-ii}) and then we
will find the
particular value of $\alpha$ and the phase of the coefficient $C_{\alpha}$
which allows the
matching of this wave packet with the wave packet (\ref{eq:p-ii}) travelling
from region II.
Next, we will find an approximate expression for the general solution
(\ref{eq:lin}) near the
horizon of region I, i.e. the surface $\xi ={\pi}/2$,  with the particular
values of $\alpha$ and the
phase of $C_{\alpha}$ calculated before. Such solution becomes exact on the
surface of the
horizon and it will be used to construct a wave packet there.

Let us start with the approximate solution of (\ref{eq:lin}) near the boundary
between regions II and I. First note that
the solution $\phi$ of the Klein-Gordon equation in the plane wave regions
(regions II or III) takes an exact WKB
form, i.e. it can be written as $\phi =C\exp{iS}$, where $C$ and $S$ are two
real functions of
the spacetime coordinates. This is directly related to the fact that the
geometrical optics
approximation is exact in the single plane wave regions, i.e. the rays of the
Klein-Gordon
solutions (the lines perpendicular to the constant phase surfaces) follow null
geodesics. Thus we
can expect that an approximate solution of the massless scalar field in the
interaction region
close to the boundaries with the plane wave regions can be obtained with the
WKB method, this is
physically related to the fact that near the boundaries the colliding plane
waves superpose
linearly. In fact from equations (\ref{eq:e'1})  and (\ref{eq:e'2}) we can see
that they
admit WKB solutions given by, see for example \cite{jon}, \bb \gamma _{{\rm
WKB}}(\xi
^*)={C\over\sqrt{Q(\xi ^*)}}\, \exp\left({\pm}i\int _{\xi ^* _0}^{\xi ^*}Q(\xi
^*)\, d\xi ^*\right),
\label{eq:wkb1} \ee  \bb \varphi _{{\rm WKB}}(\eta ^*)={C\over\sqrt{Q(\eta
^*)}}\,
\exp\left({\pm}i\int _{\eta ^* _0}^{\eta ^*}Q(\eta ^*)\, d\eta ^*\right) ,
\label{eq:wkb2} \ee where
$Q(\xi ^*)$ and ${R}(\eta ^*)$ are, \bb Q^2(\xi ^*)={k_x^2 +{{\cos}^2\xi\over
M^2(1+\sin\xi )^4}\left(\alpha +{2\sin\xi\over 1+\sin\xi}\right)},
\label{eq:q1} \ee
\bb R^2(\eta ^*)={-k_y^2 +{{\cos}^2\eta\over
M^2}\, \alpha }. \label{eq:q2} \ee
These WKB solutions can be used provided $Q(\xi ^*)$ and ${R}(\eta ^*)$ do not
vanish and they
become accurate solutions when $Q(\xi ^*)$ and ${R}(\eta ^*)$ change slowly,
i.e. when
$|dQ(\xi ^*)/d\xi ^*|<\! < Q^2(\xi ^*)$ and $|dR(\eta ^*)/d\eta ^*|<\! <
R^2(\eta ^*)$. Note that
equation (\ref{eq:e'1}) admits a WKB approximation of this kind throughout
region I. Furthermore
this approximation becomes asymptotically exact on the surface of the horizon
of region I, i.e.
$\xi ={\pi}/2$, and it is a really good approximation near the boundaries
between regions I and II
(or I and III) provided the separation constant $\alpha$ is large enough: in
fact, we will show
that to ensure the correct matching between wave packets travelling from
region II into I (or
from region III into I) with a single term in the infinite sum (\ref{eq:lin}),
it is necessary that
the terms containing
$\alpha$ in (\ref{eq:q1}) and (\ref{eq:q2}), which also contain factors
${\cos}^2\xi$ or
${\cos}^2\eta$, be dominant. Recall that when we aproach the
horizon, i.e. the surface $\xi ={\pi}/2$, on the boundary between regions I
and II (or I and III), then
both ${\cos}^2\xi$ and ${\cos}^2\eta$ go rapidly to zero. This means that
$\alpha$ must be big enough
to compense the decreasing behavior of ${\cos}^2\xi$ and ${\cos}^2\eta$.
Equation (\ref{eq:e'2}) on
the other hand admits a WKB approximation only near the boundary between
regions I and II (or I and
III) under the assumption of large $\alpha$ (this is because  of the minus
term in
(\ref{eq:q2})). Fortunately, we know its exact solution when it is written in
the form (\ref{eq:e2}),
this solution is given in terms of the {\it associated Legendre polynomials}
and that will be the
solution we will take in the regions where the WKB approximation is not valid.

Let us now evaluate the form of $\gamma _{\alpha k_x}(\xi)$ and
$\varphi _{\alpha k_y}(\eta)$ near the boundary between regions I and II, i.e.
$\{\xi =-\eta ,\;
0<\xi<{\pi}/2\}$. Although $Q(\xi ^*)$ and ${R}(\eta ^*)$, as given by
(\ref{eq:q1}) and
(\ref{eq:q2}),  are the complete terms that appear in the  WKB formulae, if we
assume that
$\alpha$ is large only the first two terms in powers of $\alpha ^{-1}$ are
relevant. With this
expansion, performing the integrals in (\ref{eq:wkb1}) and (\ref{eq:wkb2}) and
choosing the minus
sign in the  exponent of (\ref{eq:wkb1}) and the plus sign in the exponent of
(\ref{eq:wkb2}), we
have: $$ \phi (\xi ,\eta ,x,y) ={\mbox{\rm e}}^{ik_x x+ik_y y}\,\sum _{\alpha
} C_{\alpha}\,
{1\over\sqrt{\cos\xi\cos\eta}}{M\over\sqrt\alpha}\exp
\left\{i\sqrt{\alpha}\left[(\eta
-\xi)-(\eta _0-\xi _0)\right]-\right.$$ \bb -\left.i\left(
k_x^2\,\left[f(\xi)-f(\xi _0)\right]+
k_y^2\,\left[g(\eta)-g(\eta _0)\right] \right){ M^2\over 2\sqrt{\alpha}}
+O\left(\alpha ^{-3/2}\right)\right\},
\label{eq:dphi} \ee
where the functions $f$ and $g$ are
given by (\ref{eq:fg}).

Next we construct a {\it double wave packet} as
\bb {\phi}_{{\tilde k}_x,n, k_y,{\tilde k}_-,q}^{{\rm (I)}}(\xi ,\, \eta ,\,
x,\, y )={1\over\sqrt{\epsilon\delta}}\int
_{{\tilde k}_x}^{{\tilde k}_x+\epsilon}dk_x\int _{{\tilde
k}_-}^{{\tilde k}_-+\delta}dk_-\, {\mbox{\rm e}}^{-iqk_- -ink_x }
\phi _{k_xk_yk_-}^{{\rm (I)}}(\xi ,\, \eta ,\, x,\, y ),\label{eq:p-I}
\ee
and write the coefficient $C_{\alpha}$ as
and phase by
\bb C_{\alpha}\equiv\left|C_{\alpha}\right|\, {\mbox{\rm e}}^{i\theta
_{\alpha}}.\label{eq:mf}  \ee
The integrand's phase in (\ref{eq:p-I}) is then,
$${\tilde\Theta}^{({\rm I})}= -qk_- -nk_x+k_xx+k_yy+\theta _{\alpha}
-2(u-u_0)\sqrt{\alpha}-$$
\bb -\left( k_x^2\,\left[f(\xi)-f(\xi _0)\right]+
k_y^2\,\left[g(\eta)-g(\eta _0)\right]\right){ M^2\over
2\sqrt{\alpha}}+O\left(\alpha
^{-3/2}\right).\label{eq:f'-I}\ee
Now ${\tilde\Theta}^{({\rm I})}$ provides us with a natural way of matching the
wave packets in region II and the wave packets in region I. Let us assume that
the matching is
good enough if the trajectories of the wave packets in regions II and I are
joined in a
continuous and diferentiable way on the point $(u=u_0,\, v=0)$ of the boundary
between these regions. This means that
${\tilde\Theta}^{({\rm I})}$ must satisfy the three constraints, \bb \left.
{\tilde\Theta}^{({\rm
I})}\right|_{(u=u_0,\, v=0)} =\left.{\Theta}^{({\rm II})}\right|_{(u=u_0,\,
v=0)}
,\label{eq:c'1}\ee  \bb \left. {{\partial}{\tilde\Theta}^{({\rm I})}\over
{\partial}u}\right|_{(u=u_0,\, v=0)}
=\left. {{\partial}{\Theta}^{({\rm II})}\over {\partial}u}\right|_{(u=u_0,\,
v=0)} ,\label{eq:c'2} \ee
\bb \left. {{\partial}{\tilde\Theta}^{({\rm I})}\over
{\partial}v}\right|_{(u=u_0,\, v=0)} =\left. {{\partial}{\Theta}^{({\rm
II})}\over
{\partial}v}\right|_{(u=u_0,\, v=0)} .\label{eq:c'3} \ee
Using (\ref{eq:f'-I}) and (\ref{eq:f-ii}), these constraints give,
respectively,
\bb \theta _{\alpha} = -\left({k_x^2\, f(u_0)+ k_y^2\, g(u_0)}\right){L_1\over
2k_-}, \label{eq:c1}\ee
\bb 2\sqrt{\alpha} + \left({k_x^2\, {\dot f}(u_0)- k_y^2\, {\dot
g}(u_0)}\right){M^2\over 2\sqrt{\alpha}} =
\left({k_x^2\, {\dot f}(u_0)+ k_y^2\, {\dot g}(u_0)}\right){L_1\over 2k_-},
\label{eq:c2} \ee
\bb \left({k_x^2\, {\dot f}(u_0)+ k_y^2\, {\dot g}(u_0)}\right){M^2\over
2\sqrt{\alpha}} =2L_2k_-,\label{eq:c3}\ee
where ${\dot f}=df/du$, ${\dot g}=dg/du$ are, using (\ref{eq:fg}),
\bb {\dot f}(x)={(1+\sin x)^4\over \cos ^2x},\;\;\; {\dot g}(x)={1\over \cos
^2x}.\label{eq:fgpunt} \ee
Note that the constraint (\ref{eq:c1}) gives the value of the phase $\theta
_{\alpha}$ for the
coefficient $C_{\alpha}$ in  (\ref{eq:lin}), and the constraints (\ref{eq:c2})
and
(\ref{eq:c3}) define the value of $\alpha$. These two constraints are
compatible, this is not
surprising because we have already noticed that the WKB approximation is the
natural
approximation near those boundaries. The
value of $\alpha$ is,
\bb \sqrt{\alpha}\simeq\left({k_x^2\,{\dot f}(u_0)+ k_y^2\, {\dot
g}(u_0)}\right){L_1\over 4k_-},\label{eq:alfa}\ee
where we assume an expansion in even powers
of $\cos u_0$ and we recall that ${\dot f}(u_0)$ and ${\dot g}(u_0)$ are of
the order $(\cos u_0)^{-2}$. Then $\alpha$ goes like
$(\cos u_0)^{-4}$, this justifies our assumption that $\alpha\cos ^{2}\xi
>\!>1$ or $\alpha\cos
^{2}\eta >\!>1$ because both terms go like $(\cos u_0)^{-2}$ in the matching
point $(u=u_0,\,
v=0)$. Note that $(\cos u_0)^{-2}$ increases rapidly when $u_0$ differs from
zero, but even
when $u_0=0$ (\ref{eq:alfa}) with the use of (\ref{eq:r-iv}) gives
$\sqrt{\alpha}\simeq k_+L_1$ and
the wave packet (\ref{eq:p-I}) still matches with (\ref{eq:p-ii}).

It is important to note that this matching fixes the phase of the wave packet
but does not
ensure its normalization. In fact, the normalization condition fixes the
modulus of the coefficient $C_{\alpha}$, but
it is convenient to postpone the calculation of this term until we have the
form of the wave packet close to the
horizon of region I, where the wave packet normalization condition is much
simpler.

Our next step is to identify the integrand's phase defining the wave packet
(\ref{eq:p-I}) near
the horizon $\xi ={\pi}/2$. To do this we can still use the WKB approximation
(\ref{eq:wkb1}) of
equation (\ref{eq:e'1}) near the horizon, this means that we can take
asymptoticaly,
\bb \gamma _{{\rm WKB}}(\xi ^*) ={1\over\sqrt{|k_x|}}\,{\mbox{\rm
e}}^{-i|k_x|\xi ^*},\label{eq:wkb-h}\ee
where we have choosen the minus sign in the phase in order to be consistent
with (\ref{eq:dphi}). From
(\ref{eq:lin})
the scalar field $\phi$ is then given by
\bb \phi ^{{\rm (I)}}(\xi ,\eta ,x,y) ={{\mbox{\rm e}}^{ik_y y} \over
\sqrt{|k_x|}}\,
\sum _{\alpha } C_{\alpha}\, {\varphi}_{\alpha {\hat k}_{y}}(\eta )\,\times\,
\left\{\begin{array}{l}\displaystyle
{\mbox{\rm e}}^{-i|k_x|(\xi ^*-x)}\; ;\; k_x\geq 0,\\ \\
 \displaystyle {\mbox{\rm e}}^{-i|k_x|(\xi ^*+x)}\; ;\; k_x\leq
0,\end{array}\right.
\label{eq:lin-h} \ee
where $\xi ^*-x=\tilde U$, $\xi ^*+x=\tilde V$ and $U'=-2M\exp (-{\tilde
U}/2M)$, $V'=-2M\exp (-{\tilde V}/2M)$,
following equations (\ref{eq:UVtilde}), (\ref{eq:U'V'}). Notice that when
$k_x\geq 0$,
the scalar field reaches the ``roof" $V'=0$ of the horizon (strictly speaking,
the rays of
$\phi$, i.e. the lines normal to the constant phase surfaces of $\phi$, reach
the ``roof" $V'=0$),
and when $k_x\leq 0$, they reach the ``roof" $U'=0$ of the horizon. This
asymptotic solution has
been obtained  because near to the horizon $\xi ={\pi}/2$ the
dominant term in (\ref{eq:q1}) is $k_x^2$. However, it is not possible to
obtain such an asymptotic solution for
equation (\ref{eq:e'2}), because of the minus term in (\ref{eq:q2}).
Fortunately, as we have
said, we can go back to the untransformed equation (\ref{eq:e2}) which is the
equation for the
{\it associated Legendre polynomials} $P_{\alpha k_y}(\sin \eta)$. Note that
in the cyclic case
$Mk_y=m$ is an integer and we can take $\alpha =l(l+1)$ for $l$ integer, then
${\rm
e}^{ik_yy}\, P_{\alpha k_y}(\sin \eta)$ is the {\it spherical harmonic}
$Y^m_l\left(y/M,\,
{\pi}/2-\eta\right)$. The matching of the packets requires that $\alpha$ is
given by (\ref{eq:alfa}),
therefore we can take $l=l_{\alpha}\simeq\sqrt\alpha$, where $l_{\alpha}$ is
the closest integer
to $\sqrt\alpha$ for large $\alpha$. Now using the asymptotic form
(\ref{eq:lin-h}) of $\phi$ near
the horizon we can construct a double packet as in (\ref{eq:p-I}).
The integrand's phase defining the wave packet is given  by,    \bb
{\Theta}^{({\rm I})}= -qk_-
-nk_x+k_yy -k_x\left(-x{\pm}\xi ^*\right)+\theta _{\alpha},\label{eq:f-I} \ee
 where the upper sign
in (\ref{eq:f-I}) stands for $k_x\geq 0$ and the lower sign stands for
$k_x\leq 0$, and $\theta
_{\alpha}$ is the phase for the coefficient $C_{\alpha}$ given in
(\ref{eq:c1}). As we have said
the coefficients $C_{\alpha}$ are restricted to satisfy a normalization
condition, to make sure
that the scalar field $\phi$ is well normalized on the horizon. The
appropriate inner product of
any two solutions $\phi _1$, $\phi _2$ of the Klein-Gordon equation is given
on the horizon by,
see \cite{dor}: \bb\langle{\phi}_1 ,{\phi}_2\rangle =-i4M\int
\cos\eta\, d\eta\, dy\left[\int _{-\infty }^{0}\left.({\phi}_1
{\buildrel\leftrightarrow\over {\partial}}_{V'}{\phi}_{2}^{*})\right|_{U'=0}dV'
+\right. \label{eq:pe}\ee $$ +\left.\int
_{-\infty }^{0}\left.({\phi}_1 {\buildrel\leftrightarrow\over
{\partial}}_{U'}{\phi}_{2}^{*})\right|_{V'=0}dU'\right].$$
Since, as we have just seen,
$$C_l\left(l,k_x,k_y,k_-\right)\equiv C_{\alpha}\propto {\mbox{\rm
e}}^{i\theta _{\alpha}}\,\delta
\left({l-l\left(k_x,k_y,k_-\right)}\right) $$ where
$l(k_x,k_y,k_-)=l_{\alpha}$ and
$\theta _{\alpha}$ is given by (\ref{eq:c1}), the normalization condition with
respect to
(\ref{eq:pe}) leads to,  \bb C_l\left(l,k_x,k_y,k_-,\right)={ {\rm
e}^{i\theta
_{\alpha}}\over 2M}\sqrt{l_{\alpha}\over
4{\pi}k_-}\delta\left({l-l_{\alpha}}\right).\label{eq:linC} \ee

Finally with the asymptotic value (\ref{eq:f-I}) for the phase near the
horizon $(\xi ={\pi}/2)$ we can use ${\partial}\Theta ^{({\rm
I})}/{\partial}k_x ={\partial}\Theta ^{({\rm I})}/{\partial}k_- =0$ to obtain
the trajectory of the wave packet near the
horizon: \bb q=\left({k_x^2\,{ f}(u_0)+ k_y^2\, { g}(u_0)}\right){L_1\over
2k^2_-},\label{eq:t1-I} \ee
\bb x^{({\rm I})}=n{\pm}\xi ^*+{{\tilde k}_x\over{\tilde k}_-}f(u_0)
L_1.\label{eq:t2-I}  \ee
Notice that $n+({\tilde k}_x/{\tilde k}_-) L_1\, f(u_0)\equiv x^{({\rm
I})}(0)$ is the value of the coordinate $x$ when the wave
packet trajectory coming from region II crosses the boundary $v=0$ into region
I, as one can see from (\ref{eq:t1-ii}), then
(\ref{eq:t2-I}) reads,
\bb x^{({\rm I})}=x^{({\rm I})}(0){\pm}\xi ^*.\label{eq:t2'-I} \ee
This is the equation for a null geodesic in a region close to the horizon as
it can be seen by
writing it in the form
$x^{({\rm I})}\approx x^{({\rm I})}(0)\mp {\mbox{\rm ln}}({\cos\xi})$,
where $x^{({\rm I})}(0)$ is a constant for a fixed wave packet ((\ref{eq:XI*})
has been used). It
is worth noticing that this is the equation for null geodesics close to the
horizon  as one
can see in the Appendix of \cite{dor}.

Equation (\ref{eq:t2'-I}) can be written, using (\ref{eq:UVtilde}) and
(\ref{eq:U'V'}),
as
\bb U'= -2M{\mbox{\rm e}}^{x^{({\rm I})}(0)/4M}\equiv U'_0\; ;\; k_x\geq 0,
\label{eq:tU-I} \ee
\bb V'= -2M{\mbox{\rm e}}^{-x^{({\rm I})}(0)/4M}\equiv V'_0\; ;\; k_x\leq 0.
\label{eq:tV-I} \ee
That is, the trajectories of the ``in" wave packets, in the region near the
horizon, are straight
lines of $U'=U_0'={const}$ for $k_x\geq 0$ and straight lines of
$V'=V_0'={const}$ for $k_x\leq
0$.

\subsection{``Out" wave packets}

Since at the horizon the fields ${\partial}/{\partial}U'$ and
${\partial}/{\partial}V'$ become two null Killing vector fields
we can define a new complete set of ``out" modes on region I. We define
solutions with positive frequency, $\omega _+$, with
respect to the vector ${\partial}/{\partial}U'$ on the $V'=0$ ``roof" of the
horizon and positive frequency, $\omega _-$,
with respect to the vector ${\partial}/{\partial}V'$ on the $U'=0$ ``roof" of
the horizon \cite{dor,kay,unr}. These modes, close to the horizon,
are given by, \bb { u}_{ \omega _- l'm'}^{{\rm out}}(U',V',\eta ,y)
 ={1\over 2M\sqrt{ (2{\pi})2{\omega}_-}}\,
Y_{l'}^{m'}\left(y/M,{\pi}/2-\eta\right)\,{\rm
e}^{-i\omega _+ U'-i\omega _- V'}, \label{eq: outm} \ee
where the labels $\omega _-$, $m'$ and $l'$ are the three independents
separation constants of the Klein-Gordon
equation (\ref{eq:KG'}) using the asymptotic metric on the horizon of
region I, and the label $\omega _+$ is given by,
\bb 16M^2{\omega}_+{\omega}_-=l'(l'+1). \label{eq:+-} \ee
If we restrict, now, to the cyclic case the labels $l'$, $m'$  are both
discrete but $\omega _-$
is continuous. We can transform $\omega _-$ into a discrete label by
constructing a {\it single wave packet} as follows,
\bb {u}^{{\rm out}}_{{\tilde \omega}_-,n', l',m'}(U',V',\eta
,y)={1\over\sqrt{\epsilon '}}\int _{{\tilde
\omega}_-}^{{\tilde \omega}_-+\epsilon '}d\omega _- \,
{Y_{l'}^{m'}\left(y/M,{\pi}/2-\eta\right)\over 2M\sqrt{
(2{\pi})2{\omega}_-}} \,{\mbox{\rm e}}^{-in'\omega _--i\omega _+ U'-i\omega _-
V'}.\label{eq:outp} \ee
The integrand's phase is
given by,
\bb \Theta = -n'\omega _--\omega _+ U'-\omega _- V'-m'{y\over M},
\label{eq:outf} \ee
and the wave packet trajectory (i.e.. ${\partial}\Theta
/{\partial}\omega_-=0$), using relation (\ref{eq:+-}), is:
\bb {\tilde \omega}_+ U'=\left(n'+V'\right) {\tilde \omega}_-.
\label{eq:outt}\ee

Let us discuss these trajectories. On the ``roof" $U'=0$ the trajectory
(\ref{eq:outt}) is $V'=-n'={ const}\leq 0$, the coordinate $V'$ is always
negative, see (\ref{eq:U'V'}), thus
$n'$ is positive and, for a fixed wave packet, constant. This relation means
that on the ``roof" $U'=0$ the
wave  packet trajectories are straight lines with $V'=const.=-n'$.
Similarly on the ``roof" $V'=0$, the trajectories, (\ref{eq:outt}), are
$({\tilde\omega}_+/{\tilde\omega}_-) U'=n'$, this
means that they are straight lines with coordinate
$U'=const.=({\tilde\omega}_-/{\tilde\omega}_+) n'$; now the label
$n'$ is negative since the coordinate $U'$ is negative, see (\ref{eq:U'V'}).
Therefore wave packets with $n'$
positive are localized on the $U'=0$ ``roof" of the horizon and wave packets
with $n'$ negative are localized on the $V'=0$
``roof" of the horizon. These ``roofs" of the horizon are depicted in Fig. 3.

\subsection{Bogoliubov coefficients}

Using the well defined inner product on the surface of the horizon
(\ref{eq:pe}),
we can compute the
Bogoliubov transformation coefficients relating the propagated ``in" wave
packet modes from the flat region (region IV)
up to the horizon of region I and the ``out" wave packet modes defined on the
horizon. The two Bogoliubov
coefficients, defined in the usual way (i.e.. $\alpha _{{\tilde k}_x,{\tilde
k}_-,n,q,m;{\tilde\omega} _-,n',l', m'}=\langle
u^{{\mbox{\rm in}}}_{{\tilde k}_x,{\tilde k}_-,n,q,m},\, u^{{\rm
out}}_{{\tilde\omega} _-,n',l', m'}\rangle$ and  $\beta
_{{\tilde k}_x,{\tilde k}_-,n,q,m;{\tilde\omega }_-,n',l', m'}=-\langle
u^{{\mbox{\rm in}}}_{{\tilde k}_x,{\tilde k}_-,n,q,m},\, u^{{\rm
out}*}_{{\tilde\omega} _-,n',l', m'}\rangle$) are,   \bb
\left.\begin{array}{l}\alpha _{{\tilde k}_x,{\tilde
k}_-,n,q,m;{\tilde\omega }_-,n',l', m'}\\ \beta _{{\tilde k}_x,{\tilde
k}_-,n,q,m;{\tilde\omega} _-,n',l', m'}\end{array}\right\}
=\mp {1\over\sqrt{\epsilon\delta\epsilon '}} \int _{{\tilde
k}_x}^{{\tilde k}_x+\epsilon }dk_x
\int _{{\tilde
k}_-}^{{\tilde k}_-+\delta}dk_-\int _{{\tilde
\omega}_-}^{{\tilde \omega}_-+\epsilon '}d\omega _- \,{\mbox{\rm
e}}^{-ink_x-iqk_-{\pm}in'\omega _-} \label{eq:alfabetap} \ee
$$
\times {i(2M)\, ({\pm}1)^m\over \sqrt {{\pi}|k_x|\omega _-}}\,\delta _{m,\,
{\pm}m'}\,
\left|C_{l'}\right|{\mbox{\rm e}}^{i\theta _{\alpha}}\,\Gamma (1+i4M|{
k}_x|)\, {\mbox{\rm e}}^{{\pm}2{\pi}M|k_x|}
\left(2M{\omega} _{\pm}\right)^{-i4M|{ k}_x|},$$
where $l'\simeq\sqrt\alpha$ and $\theta _{\alpha}$ are given by
(\ref{eq:alfa}) and (\ref{eq:c1}) respectively. The upper signs in
(\ref{eq:alfabetap}), except for the label $\omega _{\pm}$, stand for the
$\alpha$ coefficient and the lower signs
for the $\beta$ coefficient. The label $\omega _{\pm}$  is the same for both
coefficents and
its double sign means $k_x\geq 0$ for the upper sign and $k_x\leq 0$ for the
lower sign.

In the approximation of small integration intervals $\epsilon$, $\epsilon '$
and $\delta$, and
following the stationary phase technique one can take the
non oscillating terms in the integrals as constants. For that
reason we can perform a convenient separation in the Bogoliubov coefficients
into modulus and phase as,
\bb \left.\begin{array}{l}\alpha _{{\tilde k}_x,{\tilde
k}_-,n,q,m;{\tilde\omega} _-,n',l', m'}\\
\beta _{{\tilde k}_x,{\tilde k}_-,n,q,m;{\tilde\omega }_-,n',l',
m'}\end{array}\right\}= \mp {
i(2M)\, ({\pm}1)^m\over \sqrt{{\pi}|{
k}_x|}}\, \left|\Gamma (1+i4M|{
k}_x|)\right|\, {\mbox{\rm e}}^{{\pm}2{\pi}M|k_x|} {\cal R}_{\alpha
/\beta},\label{eq:alfabetaR} \ee
where we have made use of Stirling's formula for the gamma function
\cite{gra}, i.e.
$$\Gamma (z)=z^{z-{1\over 2}}{\mbox{\rm e}}^{-z}\sqrt{2{\pi}}\left[1+{1\over
12z}+{1\over 288z^2} +
O(z^{-3})\right],$$ it can be written as
\bb\Gamma (1+iy)=
\left|\Gamma (1+iy)\right|\exp i\left({{y\over 2}\,{\mbox{\rm
ln}}({1+y^2})-y+{1\over 2}\tan ^{-1}
y}\right)\label{eq:stir'}\ee   for values $y>0$, and can be written as
\bb \Gamma (1+iy)=
\left|\Gamma (1+iy)\right|\exp ({y\,{\mbox{\rm ln}}
y-y+{\pi}/4}),\label{eq:stir}\ee
for $y>1$. We have also defined ${\cal R}_{\alpha
/\beta}$ as,
\bb {\cal R}_{\alpha /\beta} ={\delta
_{m,{\pm}m'}\over\sqrt{\epsilon\delta\epsilon '}}
\int _{{\tilde
k}_x}^{{\tilde k}_x+\epsilon }dk_x\,{\mbox{\rm e}}^{-ink_x+i\Omega}
\int _{{\tilde
k}_-}^{{\tilde k}_-+\delta}dk_-\, \left|C_{l'}\right|\,{\mbox{\rm
e}}^{-iqk_-+i\theta _{\alpha}}\label{eq:Rab}\ee
$$\int _{{\tilde
\omega}_-}^{{\tilde \omega}_-+\epsilon '}{d\omega _-\over\sqrt{\omega _-}}
\,{\mbox{\rm e}}^{{\pm}in'\omega _- - i4M|k_x|{\rm
ln}(2M\omega _{\pm})}  ,$$
where $\Omega =4M|k_x|\,{\mbox{\rm ln}}\left(4M|k_x|\right)-4M|k_x|+{\pi}/4$,
it comes from Stirling's
formula (\ref{eq:stir}), where we have assumed $4M|k_x|>1$ for simplicity.
This assumption does not
afect our results because the purpose is to relate a single ``in" wave packet
with a fix
label ${\tilde k}_x$ to a single ``out" wave packet with a fix label ${\tilde
\omega}_- $. Of
course we can follow the same steps for the case $4M|k_x|>0$ using the more
accurate version of the
Stirling's formula (\ref{eq:stir'}). We have kept the terms
$\left|C_{l'}\right|$ and $(\omega
_-)^{-1/2}$ in the integrands for convergence.

The terms, ${\cal R}_{\alpha }$ and ${\cal R}_{\beta}$ act like localizing
terms in the sense that the three
integrals, of which they consist, have a well localized peak. The integrand's
phase of
 ${\cal R}_{\alpha }$ and ${\cal R}_{\beta}$ (${\Theta}_{\alpha }$ and
${\Theta}_{\beta}$,
respectively) are
\bb \Theta _{\alpha /\beta}=-nk_x-qk_-{\pm}n'\omega _-+\Omega +\theta
_{\alpha} -
4M|k_x|{\mbox{\rm ln}}(2M\omega _{\pm}), \label{eq:fab}\ee
and the peaks of  ${\cal R}_{\alpha }$ and ${\cal R}_{\beta}$ are in the
spacetime points which satisfy
${\partial}\Theta _{\alpha /\beta}/{\partial}k_x={\partial}\Theta _{\alpha
/\beta}/{\partial}k_-={\partial}\Theta _{\alpha /\beta}/{\partial}\omega _-=0$.
Solving these
three equations for $\Theta _{\alpha}$ we find, respectively,
\bb n={{\partial}\theta _{\alpha}\over {\partial}k_x}{\pm} 4M{\mbox{\rm
ln}}\left({2|{\tilde k}_x|\over{\tilde\omega} _{\pm}}\right),\label{eq:alpha1}
\ee
\bb q={{\partial}\theta _{\alpha}\over {\partial}k_-},\label{eq:alpha2} \ee
\bb n'=-4M{{\tilde k}_x\over{\tilde\omega} _-},\label{eq:alpha3} \ee
where the upper sign in (\ref{eq:alpha1}) stands for $k_x\geq 0$ and the lower
sign for $k_x\leq 0$. Similarly, from the
equations for $\Theta _{\beta}$, we obtain:
\bb {\bar n}={{\partial}\theta _{\alpha}\over {\partial}k_x}{\pm} 4M{\mbox{\rm
ln}}\left({2|{\tilde{\bar k}}_x|\over{\tilde\omega}
_{\pm}}\right),\label{eq:beta1}
\ee \bb {\bar q}={{\partial}\theta _{\alpha}\over
{\partial}k_-},\label{eq:beta2} \ee
\bb n'=4M{{\tilde{\bar k}}_x\over{\tilde\omega} _-},\label{eq:beta3} \ee
where the two first equations are functionally the same as (\ref{eq:alpha1})
and (\ref{eq:alpha2}) but with a bar over
the ``in" labels, i.e. ${\tilde{\bar k}}_x$, ${\bar n}$, ${\bar q}$, since
these three equations come from the $\beta$
coefficient which relates an ``out" wave packet with an ``in" anti wave packet
(i.e. a wave packet constructed by the
superposition of monocromatic modes of negative frequency). On the other hand
(\ref{eq:alpha1}),
(\ref{eq:alpha2}) and (\ref{eq:alpha3}) come from an $\alpha$ coefficient and
so relate an
``out" wave packet with an ``in" wave packet. In fact,  notice that equations
(\ref{eq:alpha1}),
(\ref{eq:alpha2}) and (\ref{eq:alpha3}) give a relation between the labels of
the ``in" wave packets
(i.e. ${\tilde k}_x$, ${\tilde k}_y$, ${\tilde k}_-$, $n$ and $q$) and the
labels of the ``out" wave
packets (i.e. $\omega _-$, $n'$, $l'$ and $m'$) and equations
(\ref{eq:beta1}), (\ref{eq:beta2}) and
(\ref{eq:beta3}) give a relation between the labels of the ``in" anti wave
packets (i.e.
${\tilde{\bar k}}_x$, ${\tilde{\bar k}}_y$, ${\tilde{\bar k}}_-$, ${\bar n}$
and ${\bar q}$) and the
labels of the ``out" wave packets (i.e. $\omega _-$, $n'$, $l'$ and $m'$).
When the previous equations
are satisfied the terms ${\cal R}_{\alpha }$ and ${\cal R}_{\beta}$ have a
peak, otherwise
they roughly vanish.

Let us now extract some more information from equations
(\ref{eq:alpha1})-(\ref{eq:beta3}).
For instance, equation (\ref{eq:alpha1}) (or (\ref{eq:beta1})) can be written
as,
\bb |k_x|={1\over 2}\omega _{\pm}\,{\mbox{\rm e}}^{{\pm}x^{({\rm
I})}(0)/4M}=\left\{\begin{array}{l} -\omega_+U'_0/(4M);\;\;
k_x\geq 0,\\-\omega_-V'_0/(4M);\;\;
k_x\leq 0,\end{array}\right. \label{eq:beta1'} \ee
where we have used (\ref{eq:c1}), (\ref{eq:tU-I}), (\ref{eq:tV-I}) and the
fact that $n+({\tilde k}_x/{\tilde k}_-)L_1f(u_0)=
x^{({\rm I})}(0)$ is the value of the coordinate $x$ when the ``in" wave
packet coming from region II crosses the boundary
$\{ v=0,\; 0\leq u<{\pi}/2\}$ into region I (see (\ref{eq:t1-ii})). Note that
$U'_0$ or $V'_0$ are the
coordinates on the horizon reached for the wave packets. From
(\ref{eq:beta1'}) equations (\ref{eq:alpha1}) and (\ref{eq:beta1})
can be seen as {\it redshift formulae}, because they relate the energy of an
``out" wave packet (i.e.
$\omega _-$) with the energies of an ``in" wave packet and an ``in" anti wave
packet, respectively,
(the energy label for the ``in" wave packets in the region near the horizon is
$|k_x|$). Note also
from (\ref{eq:beta1'}), that the {\it  redshift coefficient} is given by the
value of the coordinate
$U$ or $V$ on the horizon reached by the ``in" wave or anti wave packets and
by the surface gravity of the horizon, i.e.
$(4M)^{-1}$. Equations (\ref{eq:alpha3}) and (\ref{eq:beta3}) are {\it
position formulae} because they give information on the
trajectories of the ``in" wave and anti wave packets which are related to a
given ``out" wave packet. In fact, the
label $n'$ appearing in these formulae can be written, in analogy to
(\ref{eq:beta1'}), as  \bb
|n'|=\left\{\begin{array}{l}\displaystyle
-{{\tilde\omega}_+\over{\tilde\omega}_-}\, U_0\; ;\;
k_x\geq 0, \\ -V_0\; ;\;\;\;\; k_x\leq 0.\end{array}\right. \label{eq:n'}\ee
Recall that the label
$n'$ of the ``out" wave packet is related to its position on the horizon, see
(\ref{eq:outt}), when
$n'\geq 0$ the ``out" wave packet is localized on the $U'=0$ ``roof" at
coordinate $V'\equiv V'_{\rm
max}=-n'={ const}$, and when $n'\leq 0$ it is localized on the $V'=0$ ``roof"
at coordinate $U'\equiv
U'_{\rm max}=({\tilde\omega}_-/{\tilde\omega}_+) n'={const}$. From
(\ref{eq:alpha3}) we see that when
$n'\geq 0$ then ${\tilde k}_x\leq 0$, and viceversa, and from
(\ref{eq:beta3}), that when $n'\geq 0$
then ${\tilde{\bar k}}_x\geq 0$, and viceversa. Equation (\ref{eq:alpha2}) (or
(\ref{eq:beta2})), with the use of (\ref{eq:c1}), is the same as
(\ref{eq:t1-I}) and it does not give
us any additional information.

Putting all this together we can give the following interpretation.
An ``out" wave packet $({\tilde\omega}_-, n')$, which is localized at the
coordinate $V'_{\rm max}$ on the ``roof" $U'=0$
($U'_{\rm max}$ on the ``roof" $V'=0$) of the horizon, is related to an ``in"
wave packet with
momentum along the $x$ axis ${\tilde k}_x={\tilde\omega}_- V'_{\rm max}/4M$
(${\tilde
k}_x=-{\tilde\omega}_+ U'_{\rm max}/4M$), which reaches the horizon at
coordinate $V_0'=V'_{\rm max}$ on the ``roof" $U'=0$
($U_0'=U'_{\rm max}$ on the ``roof" $V'=0$), and to an ``in" anti wave packet
with the same momentum
along the $x$ axis, but with opposite sign,  ${\tilde{\bar
k}}_x=-{\tilde\omega}_- V'_{\rm max}/4M$
(${\tilde{\bar k}}_x={\tilde\omega}_+ U'_{\rm max}/4M$), which reaches the
horizon at coordinate
${\bar U}_0'=({{\tilde\omega}_-/{\tilde\omega}_+})V'_{\rm max}$ on the ``roof"
$V'=0$
(${\bar V}_0'=({{\tilde\omega}_+/{\tilde\omega}_-})U'_{\rm max}$ on the
``roof" $U'=0$).

\subsection{Particle creation}

Following the formalism of quantum field theory on curved spacetime,
spontaneous particle creation is directly related to the
${\beta}$ Bogoliubov coefficient. In fact, the number of ``out" particles in a
given wave packet mode with labels
$({\tilde\omega}_-,n',l',m')$ (i.e. the number of quanta in the wave packet
mode
$({\tilde\omega}_-,n',l',m')$)
in the ``in" vacuum is given by the sum over the ``in" labels of the squared
modulus of the  ${\beta}$ coefficient \cite{bir},
(\ref{eq:alfabetaR}), that is,
\bb \langle 0,\,{\mbox{\rm in}}|N^{\rm
out}_{{\tilde\omega}_-,n',l',m'}|0,\,{\mbox{\rm in}} \rangle =\sum _{{\tilde
k}_x,{\tilde
k}_-,m,n,q}\,\left|\beta _{ {\tilde k}_x,{\tilde
k}_-,m,n,q;\,{\tilde\omega}_-,n',l',m' }\right|^2. \label{eq:number} \ee
Here $N^{\rm out}_{{\tilde\omega}_-,n',l',m'}$ is the number operator of
``out" particles defined in the standard way as
$N^{\rm out}_{{\tilde\omega}_-,n',l',m'}=a^{\dagger\rm
out}_{{\tilde\omega}_-,n',l',m'}a^{\rm
out}_{{\tilde\omega}_-,n',l',m'}$, where $a^{\dagger\rm
out}_{{\tilde\omega}_-,n',l',m'}$ and $a^{\rm
out}_{{\tilde\omega}_-,n',l',m'}$ are the ``out"  wave packet creation and
annihilation operators
respectively. With these operators we can write the field operator $\phi$, as
a combination of ``out"
wave and anti wave packets, i.e. $$\phi (x)
=\sum_{{\tilde\omega}_-,n',l',m'}a^{\rm
out}_{{\tilde\omega}_-,n',l',m'} u^{\rm out}_{{\tilde\omega}_-,n',l',m'}({
x})+a^{\dagger\rm
out}_{{\tilde\omega}_-,n',l',m'}u^{*\rm out}_{{\tilde\omega}_-,n',l',m'}({
x}).$$
To evaluate the sum
in (\ref{eq:number}) it is worth noticing that the ${\cal R}_{\beta}$
coefficient in
(\ref{eq:alfabetaR}), given by (\ref{eq:Rab}), satisfy the following two
equalities:
\bb \sum
_{{\tilde\omega}_-,n',l',m'}\,\left|{\cal R}_{\beta}\right|^2={1\over
4(2M)^3}, \label{eq:prop1}\ee
\bb \sum _{{\tilde k}_x,{\tilde k}_-,m,n,q}\,\left|{\cal
R}_{\beta}\right|^2={1\over 4(2M)^3}.
\label{eq:prop2} \ee
These equalities follow from the general property (\ref{eq:wpprop}) of
wave packets and from the form of $C_{l'}$ given by (\ref{eq:linC}). Note
also, in order to make sense of the square of
$C_{l'}$ according to (\ref{eq:linC}) we consider the product of a Dirac's
delta with a Kronecker's delta.

These two equalities
and the fact that the ${\cal R}_{\beta}$ term has a peak when the relations
(\ref{eq:beta1})-(\ref{eq:beta3}), between the ``in" labels (i.e. $k_x$,
$k_y$, $k_-$, $n$, $q$) and
the ``out" labels (i.e. ${\omega}_-$, $n'$, $l'$, $m'$) are satisfied, allow
us to approximately write
\bb {\cal R}_{\beta}\simeq{{1\over 2(2M)^{3/2}}}\,\delta _{{\tilde k}_x,\,
n'{\tilde\omega}_-(4M)^{-1}}\,\delta_{n,\, {\bar
n}({\tilde\omega}_-,n',l',m')}\,\delta_{k_-,\,
k_-({\tilde\omega}_-,n',l',m')}\,\delta_{q,\,
q({\tilde\omega}_-,n',l',m')}\,\delta_{m,-m'},\label{eq:Raprox}\ee    where
$n={\bar n}({\tilde\omega}_-,n',l',m')$ is given by (\ref{eq:beta1}),
$q=q({\tilde\omega}_-,n',l',m')$ by
(\ref{eq:beta2}), and $\delta_{k_-,\, k_-({\tilde\omega}_-,n',l',m')}$ is
given by
$\delta_{l',l_{\alpha}}$ which appears from the squared modulus of
(\ref{eq:linC}). Then the number
of ``out" particles created in the ``out" packet mode
$({\tilde\omega}_-,n',l',m')$, equation
(\ref{eq:number}), is simply,  \bb \langle 0,\,{\mbox{\rm in}}|N^{\rm
out}_{{\tilde\omega}_-,n',l',m'}|0,\,{\mbox{\rm in}}\rangle\equiv N^{\rm
out}_{{\tilde\omega}_-,n'}={1\over {\rm
e}^{8{\pi}M{\tilde\omega}_-\left(|n'|/4M\right)}-1} ,\label{eq:espectre}\ee
which can be interpreted as
a thermal spectrum for each fixed value of the label $n'$, with a temperature,
 \bb T={1\over
8{\pi}M\left(|n'|/4M\right)}. \label{eq:temperatura} \ee This spectrum depends
on the dimensionless label
$n'/4M$, i.e. on the trajectory of the wave packet, but it is thermal for all
wave packet modes with
the same trajectory. This is quite different from the black hole case
\cite{haw} where the
temperature is independent of the packet trajectory \cite{mul} and depends
only on the surface gravity,
$\kappa =(4M)^{-1}$. In ref. \cite{dor} we discuss how the black hole case can
be seen in
some sense as the time reversal of the colliding wave case. The physical
interpretation of this
$n'$-dependent temperature follows from the fact that we are computing the
particles produced on the
wave packet mode $({\tilde\omega}_-,n',l',m')$, which is localized on the
horizon by $n'$, given by
(\ref{eq:outt}), so that these particles may be ``localized" in the same
position on the horizon. Note
that when $n'\simeq 0$, i.e. near the bifurcation point $U'=V'=0$ of the
horizon, the temperature is
higher.

This spectrum is in agreement with the spectrum of
particles created on the monocromatic modes (\ref{eq:CSPT}), i.e. particles
with a well defined momentum but not localized in
space, because the one particle Fock space can be decomposed on a basis given
in terms of the monocromatic labels,
$|\omega \rangle$, or, alternatively, on a basis given in terms of the wave
packet labels, $|{\tilde\omega}, n' \rangle$. The
space is the same but, of course, the particle interpretation is different.
For the monocromatic modes, discused in ref. \cite{dor}, the spectrum of
particles created (\ref{eq:CSPT}) is inversely proportional to the energy of
the ``out" modes (i.e. $\omega _-$), with a
proportionality factor $(8{\pi}M)^{-1}$.
The relation between the number operators $N^{\rm out}_{{\omega}_-}$ and
$N^{\rm out}_{{\tilde\omega}_-,n'}$ is given in
terms of the Bogoliubov coefficients relating the monocromatic modes and the
wave packet modes. Such transformation has no
$\beta$ Bogoliubov coefficient because the positive frequency wave packets are
constructed with positive frequency
monocromatic modes only. The $\alpha$ Bogoliubov coefficient is
$$\alpha _{\omega _- ;\omega _-,n'}={{\mbox{\rm
e}}^{-in'\omega}\over\sqrt{\epsilon '}}\left[\theta (\omega _-
-{\tilde\omega}_-)-\theta(\omega _-
-\{{\tilde\omega}_-+\epsilon '\})\right],$$
where, here, $\theta (x)$ is the usual Heaviside step function, thus the
relation
between the monocromatic and wave packet annihilation operators is given by
$$a^{\rm out}_{\omega _-}=\sum_{{\tilde\omega} _-, n'}\alpha ^*_{\omega _-
;\omega _-,n'}
a^{\rm out}_{{\tilde\omega} _-,n'}={1\over\sqrt{\epsilon
'}}\sum_{n'}{\mbox{\rm e}}^{-in'\omega _-}
a^{\rm out}_{{\omega} _-,n'}.$$
Then the monocromatic number operator can be
written as
$$N^{\rm out}_{{\omega}_-}=a^{\dagger{\rm out}}_{{\omega} _-}a^{\rm
out}_{{\omega}
_-}={1\over\epsilon '}\sum_{n',{\tilde n}' } {\mbox{\rm e}}^{-i(n'-{\tilde
n}')\omega _-}a^{\dagger{\rm
out}}_{{\omega} _-,n'}a^{\rm out}_{{\omega} _-,{\tilde n}'},$$ and we can
approximate
$$\langle 0,\,{\mbox{\rm in}}|N^{\rm out}_{{\omega}_-}
|0,\,{\mbox{\rm in}}\rangle \simeq{1\over\epsilon '} \sum_{n'}
\langle 0,\,{\mbox{\rm in}}|a^{\dagger{\rm out}}_{{\omega} _-,n'}
a^{\rm out}_{{\omega} _-,n'}|0,\, {\mbox{\rm
in}}\rangle= {1\over\epsilon '} \sum_{n'}N^{\rm out}_{{\omega} _-,n'},$$ as
the phase term
${\exp}[{-i(n'-{\tilde n}')\omega _-}]$ is fast oscillating (except when
$n'={\tilde n}'$), because
$n'-{\tilde n}'=(l-{\tilde l})2{\pi}/\epsilon '$ with $l$ and ${\tilde l}$
integers and $\epsilon '$ a
small positive parameter. Then
\bb \langle 0,\,{\mbox{\rm in}}|N^{\rm out}_{{\omega}_-}
|0,\,{\mbox{\rm in}}\rangle\simeq {2M\over {\pi}}\, {1\over
8{\pi}M{\omega}_-}\,\int _0^\infty {dk\over
{\mbox{\rm e}}^{k}-1},\label{eq:limit}  \ee
where we have assumed $8{\pi}M\omega _-<\! < 1$ in order to
approximate $k\equiv (8{\pi}M{\omega}_-) n'/ 4M$ as a continuous dimensionless
variable. The
inverse proportionality in $\omega _-$ and the logarithmic divergence, in
(\ref{eq:CSPT}),  have been recovered. The extra
factor $(2M)^2$ in (\ref{eq:CSPT}) is due to the fact that two monocromatic
labels in ref. \cite{dor} were, in fact, discrete
and this factor guarantees the correct normalizations.

\vskip 1.25 truecm

{\Large{\bf Acknowledgements}}

\vskip 0.5 truecm

\noindent
We are grateful to J. A. Audretsch and R. M\"uler for helpful discussions.
This work has been partially supported by a CICYT
research project number AEN93-0474.

\vskip 0.7 truecm
{\Large{\bf Figure Captions}}
\vskip 0.5 truecm

{\bf Fig.1}
\vskip 0.5 truecm

This figure shows a projection of the colliding plane wave spacetime in
the $(u,v)$ plane. One can see four regions: region IV is the flat region
before the waves collide, region II and III are the plane wave regions and
region I is the interaction region. The interaction starts at $u=v=0$,
the lines $u=0,\, v<0$ and $v=0,\, u<0$ are the boundaries of region IV
with regions II and III respectively, and $v=0,\, 0\leq u<{\pi}/2$ and $u=0,\,
0\leq v<{\pi}/2$ are the boundaries of region I with regions II and III
respectively. At $u+v={\pi}/2$ a Killing-Cauchy horizon is formed which one can
see as a coordinate singularity of the metric in region I.

\vskip 0.5 truecm
{\bf Fig.2}
\vskip 0.5 truecm

In this figure we represent the coordinates $(\xi ,x)$ in terms of the
Kruskal-Szekeres like coordinates $(U',V')$ in the interaction region. The
lines $\xi =constant$ are
hyperbolae and  the $x=constant$ are straight lines crossing the origin
$U'=V'=0$. The Cauchy horizon is $\{U'=0,\, V'<0\}\cup\{V'=0,\, U'<0\}$ which
corresponds to the limit of the hyperbolae as $\xi\rightarrow 0$. The
``roof" $U'=0$ corresponds to $x\rightarrow -\infty$ and the "roof" $V'=0$ to
$x\rightarrow\infty$.

\vskip 0.5 truecm
{\bf Fig.3}
\vskip 0.5 truecm

This is a 3-dimensional plot of the interaction region (region I) in which all
its
boundaries are shown, using non singular Kruskal-Szekeres like coordinates.
The surface
${\cal S}_3$ and the lines  ${\cal M}_1$, ${\cal M}_2$, ${\cal M}'_1$ and
${\cal M}'_2$, are
the boundaries of region I and the two plane wave regions II and III, the
points $\cal P$
and ${\cal P}'$ correspond to $U'=0,\, V'=0,\, \eta =-{\pi}/2$ and to $U'=0,\,
V'=0,\, \eta
={\pi}/2$, respectively, and they are {\it folding singularities}
\cite{dor,hay,mat}. The
Cauchy horizon is the ``roof" $\{U'=0,\, V'<0\}\cup\{V'=0,\, U'<0\}$ and
region I is
enclosed between the surface ${\cal S}_3$ and the ``roof".


\begin{thebibliography}{99}


\bibitem{kah} K.\ Kahn and R.\ Penrose, {\sl Nature (London)} {\bf 229} (1971)
185.

\bibitem{gri} J.\ B.\ Griffiths, {\sl Colliding waves in general
relativity}, Clarendon Press, Oxford, (1991)

\bibitem{sze}  P.\ Szekeres, {\sl Nature (London)} {\bf 228} (1970) 1183.

\bibitem{eat} P.\ D.\ D'Eath, in {\sl Sources of gravitational radiation},
edited by L.\ Smarr (Cambridge University Press, Cambridge, England, 1979)

\bibitem{fer-1} V.\ Ferrari, P.\ Pendenza and G.\ Veneziano, {\sl Gen. Rel.
Grav.} {\bf 20} (1980) 1185.

\bibitem{yur-1} U.\ Yurtsever, {\sl Phys. Rev.} {\bf D40} (1989) 360.

\bibitem{fin} D.\ Garfinkle, {\sl Phys. Rev.} {\bf D41} (1989) 1112.

\bibitem{vach} D.\ Garfinkle and T.\ Vachaspati, {\sl Phys. Rev.} {\bf D42}
(1990) 1960.

\bibitem{dor} M.\ Dorca and E.\ Verdaguer, {\sl Nucl.
Phys.} {\bf B403} (1993) 770.

\bibitem{hay} S.\ A.\ Hayward, {\sl Class. Quantum Grav.} {\bf 6} (1989) 1021.


\bibitem{cha-1} S.\ Chandrasekhar and B.\ Xanthopoulos, {\sl Proc. R. Soc.
(London)} {\bf A408} (1986) 175.

\bibitem{cha-2} S.\ Chandrasekhar and B.\ Xanthopoulos, {\sl Proc. R. Soc.
(London)} {\bf A410} (1987) 311.

\bibitem{kay} B.\ S.\ Kay and R.\ M.\ Wald, {\sl Phys. Rep.} {\bf
207} (1991) 49.

\bibitem{haw} S.W. Hawking, {\sl Commun. Math. Phys.} {\bf 43} (1975) 199.

\bibitem{yur-2} U.\ Yurtsever, {\sl Phys. Rev.} {\bf D38} (1988) 1706.

\bibitem{wal-2} R.\ M.\ Wald {\sl Phys. Rev.} {\bf D13} (1976) 3176.

\bibitem{mul} J.\ A.\ Audretsch and R.\ M\"uller, {\sl Phys. Rev.} {\bf D45}
(1992) 513.

\bibitem{mul'} J.\ A.\ Audretsch and R.\ M\"uller, {\sl University of
Konstanz, preprint} (1993).


\bibitem{eck} C.\ Eckart, {\sl Rev. Mod. Phys.} {\bf 20} (1948) 399.

\bibitem{jon} J.\ Mathews and R.\ L.\ Walker, {\sl Mathematical Methods of
Physics}
(Addison-Wesley, California, 1964).

\bibitem{unr} W.\ G.\ Unruh, {\sl Phys. Rev.} {\bf D14} (1976) 870.

\bibitem{gra} I.\ S.\ Gradshteyn and I.\ M.\ Ryzhik, {\sl Table of integrals,
series and products} (Academic Press,
New York, 1980).

\bibitem{bir} N.\ D.\ Birrell and P.\ C.\ W.\ Davies, {\sl Quantum fields in
curved space} (Cambridge University Press, Cambridge, England, 1982).

\bibitem{mat} R.\ A.\ Matzner and F. J.\ Tipler, {\sl Phys. Rev.} {\bf D29}
(1984) 1575.


\end{thebibliography}
\end{document}